\documentclass[11pt]{article}

\newcommand{\hf}{\frac{1}{2}}

\newcommand{\xn}{x_{n}}

\newcommand{\e}{e^{i kX(z)}}

\newcommand{\kom}{ k_{0}^{\mu}}                                      
\newcommand{\ki}{ k_{1}}
\newcommand{\yn}{ Y_{n}}                                             
\newcommand{\ym}{ Y_{m}} 
\newcommand{\kn}{ k_{n}}

\newcommand{\km}{ k_{m}}

\newcommand{\ko}{ k_{0}}                                             
                                             
\newcommand{\yim}{ Y_{1}^{\mu}}

\newcommand{\kon}{ k_{0}^{\nu}}

\newcommand{\gvkt}{ e^{i\sum _{n \ge 0 }k_{n}\tY_{n}(z)}}

\newcommand{\dsn}{\frac{\partial }{\partial x_{n}}}

\newcommand{\p}{\partial}                                           
\newcommand{\pb}{\bar \partial}

\newcommand{\al}{\alpha }                                             
\newcommand{\aln}{\alpha _{n}} 
\newcommand{\tY}{\tilde Y}

\newcommand{\la}{\mbox{$ \lambda $}} 
\newcommand{\be}{\begin{equation}}
\newcommand{\br}{\begin{eqnarray}}
\newcommand{\ee}{\end{equation}} 
\newcommand{\er}{\end{eqnarray}}

\begin{document}
\title{
\hfill\parbox{4cm}{\normalsize IMSC/2007/12/17 \\ 
                               }\\
\vspace{2cm}
Gauge Invariant Action for the Open Bosonic String.
\author{B. Sathiapalan\\ {\em Institute of Mathematical Sciences}\\
{\em Taramani}\\{\em Chennai, India 600113}\\ bala@imsc.res.in}}
\maketitle       

\begin{abstract} 
The issue of space time gauge invariance for the bosonic string has
been earlier addressed using the loop variable formalism. 
 In this paper the question of obtaining a gauge invariant
action for the open bosonic string is discussed. The derivative w.r.t $ln ~a~$ (where $a$ is a
world sheet cutoff) of  the partition function - which is first normalized by
dividing by the integral of the two point function of a marginal operator - 
 is a
candidate for the action. Applied to the zero-momentum  tachyon it
gives a tachyon potential that is  similar to those that have been obtained 
using Witten's background independent formalism.  This procedure 
is easily made gauge invariant in the loop variable formalism by replacing 
$ln ~a$ by $\Sigma$ which is the generalization of the Liouville
mode that occurs in this formalism. We also describe a method of resumming
the Taylor expansion that is done in the loop variable
formalism. This allows one to see the pole structure of string
amplitudes that would not be visible in the original loop variable formalism.

\end{abstract}
\newpage

\section{Introduction}

A gauge invariant and (manifestly) background independent formalism for string theory has
been the subject of a lot of investigation \cite{WiI,WiII,LW,ShI,ShII}.  The sigma model approach 
(\cite{L} - \cite{T}) is manifestly background
independent - in the sense that it does not involve as a starting point, a conformal field theory.
One can turn on any background and calculate either the beta functions or the equations of
motion.  However as it stands,
the beta functions or the equations of motion that one obtains in the sigma model approach
are not gauge invariant. String theory has, in addition to the usual invariances associated
with massless fields, invariances associated with massive fields also. These need to be reflected
in the equations. This is probably a prerequisite if one is to acquire a deeper understanding
of string theory - that goes beyond computational techniques.   One elegant proposal is
 \cite{WiI,WiII,LW}
based on the Batalin-Vilkovisky formalism. The action  is (at least formally)
 gauge invariant \footnote{ Some issues about removing the cutoff dependence 
 in this approach have been discussed in \cite{LW}}. This formalism has has been discussed further
and used
to derive the tachyon potential in \cite{ShI,ShII,KMM}. 

The loop variable approach \cite{BSLV}  is also based on the sigma model and is also
 gauge invariant.  We will not describe it here in any detail (see \cite{BSREV} for a review)
except to say that it is motivated by some speculations \cite{BSLV}
about the underlying principles of string theory. In earlier papers gauge invariant equations of motion
were derived, first for the open bosonic string, and subsequently the introduction of Chan Paton factors
was worked out as was the closed string. Keeping the aim of background independence, 
the formalism was also generalized to curved space and gauge invariant and generally covariant 
equations for  massive higher spin modes of the open string  in curved space were worked out. 

However, what is obtained by this procedure is actually the gauge invariant 
generalization of the beta functions. In fact in general these equations cannot be derived from an action. 
This issue was addressed for a free (open string) higher spin massive mode in AdS space. For this particular
case it was shown that relatively minor modifications of the equations could be made that
ensured that they could be obtained from an action, and the action was also worked out. However the general
problem remains.

In this paper we give a tentative prescription for obtaining directly a gauge invariant action in the loop variable
formalism. In the case of the tachyon, which is easiest to handle, it gives results similar to those
that have been obtained in other approaches \cite{WiI,WiII,ShI,ShII,KMM}.   We also address another issue
that is present in the loop variable formalism. Although the interacting equations are gauge invariant
they are not in a convenient form. This is because a Taylor expansion is performed that
makes obscure the pole structure of the string (Veneziano-like) amplitudes. We show here that 
it is possible to do a resummation that makes the pole structure manifest, without sacrificing
gauge invariance.

This paper is organized as follows: In Section 2 we give the prescription for
the Action in the gauge fixed case. In Section 3 we apply it to the case of the tachyon. In Section 4 we give
the gauge invariant generalization.  In Section 5 a method of resummation is described. Section 6 contains
some conclusions.

\section{Action}
\setcounter{equation}{0}

We consider the disk (or equivalently the upper half plane (UHP)) partition function $Z$ in the presence of
background fields. This corresponds to tree diagrams in the space time field theory.
This can be represented as
\be
Z= \langle e^{-\int {dz \over a} \sum _i g^i  a^{\Delta _i} O_i(z)}\rangle
\ee
Here $\Delta _i$ is the engineering (mass) dimension of the operator $O_i$, and 
appropriate powers of the short distance cutoff '$a$' have been introduced
so that the couplings $g^i$ are  dimensionless. 
The renormalization group (RG) equations require that $g^i$ change
in such a way that the partition function is independent of the cutoff $a$.
\[
{d\over d~ln~a} Z =0
\]
We are assuming that  this cutoff dependence can be entirely absorbed into
the various coupling constants: $g^i (a)$. So a change in $a$ is compensated by
changing all the $g^i$ as specified by the corresponding $\beta$-function.  
If this is to be true for any $a$, there must necessarily be an infinite number of operators.
If it is to be true only in the continuum limit, $a\rightarrow 0$,  then we can get away with
a finite number of renormalizable operators.
In the  present context we are considering all possible operators - since the world sheet
theory, involving massive background fields, will have higher dimension operators. Thus we are
actually dealing with the exact renormalization group (ERG). \footnote{Some aspects of this were dealt
with in \cite{BSERG}.}
Thus
\be	\label{beta}
{dZ\over d~ln~a } =0 = {\p Z \over \p ln~a} + {\p g^i \over \p ~ln~a} {\p Z\over \p g^i} 
={\p Z \over \p ln~a} - \beta ^i  {\p Z\over \p g^i}
\ee 
 (Note that $g(a)$ are
what are conventionally called "bare" couplings, and in this convention
a relevant bare coupling gets smaller as the short distance cutoff is made smaller ($a\rightarrow 0$),
i.e. the beta function is negative.) To lowest order we simply get
\[
\beta ^i = (\Delta _i-1)g^i
\]
In the RG approach the $\beta$-functions are proportional to the equations
of motion. In this section a prescription for the action will be motivated
using  the tachyon where the
issue of gauge invariance will not arise. Thus 
\be
Z = \langle e^{-\int _{\p \Gamma} {dz \over a} \phi (X(z))}\rangle
\ee  
We can write $\phi (X(z)) = \int dk~\phi (k)\e$ and then the operators $O_i$ are $\e$ with
the space-time momentum, $k$, playing the role of the index $i$.

The expectation value is calculated using the Polyakov measure
\[
\int {\cal D} X e ^{- {1\over \al '}\int _{\Gamma} d^2z \p X \pb X }
\]

We are working with a Euclidean metric on the world sheet. $\Gamma$ is
the upper half plane (UHP) because we are interested in the open
string and $\p \Gamma$ is the real axis. With this
normalization $\langle X(z) X(w) \rangle = - {\al '\over 2\pi} (ln ~ |z-w|
+ ln ~ |z-\bar w|)$ for the upper half plane with Neumann boundary
conditions on the real axis. Further, for a tachyon vertex operator
on the real axis at $z=x$, $e^{ik.X(x)} = :e^{ik.X(x)}: e^{{\al '
    k^2 \over 2\pi }ln ~ a}$. Thus if we set $\al '=\pi$, $k^2=2$ 
ensures that the operator is marginal and this is
the mass shell condition for the open string tachyon. 

Let us evaluate $Z$ in powers of $\phi$.
\be
Z~=~\langle 1 \rangle ~+~\langle-\int _{-R}^{+R} {dz \over a} \phi
(X(z))    \rangle  ~+~\langle \int _{-R+a}^{+R}  {dz_1 \over a}  
\phi (X(z_1))\int _{-R}^{z_1-a} {dz_2 \over a} \phi (X(z_2))  \rangle +...
\ee

Assume for the moment that $\phi$ has non-zero space-time momentum so
that it is of the form $\int dk~\phi (k) \e$ with $\phi (0) =0$.
Then the linear term vanishes by momentum conservation. The quadratic
term is 
\[
\int dp ~\int dk~ \int _{-R+a}^{R} {dz \over a} ~\int _{-R} ^{z-a}
{du \over a} (z-u)^{k.p}\phi (k)  \phi (p) \delta (k+p)a^{k^2 +p^2\over 2}
 \]
\be
=~\int dk \int _{-R+a}^{R} {dz \over a} ~\int   _{-R} ^{z-a}{du \over a}
(z-u)^{-k^2}a^{k^2}
\ee
 
The kinetic term for $\phi$ has to come from this. We would like to
obtain it by the action of the renormalization group (RG) operation of 
${d\over d~ln~a }$. 
 This enables us to replace it in the loop
variable formalism by the operation of ${d\over d \Sigma}|_{\Sigma
  =0}$,which gives a gauge invariant object. (This is seen as follows:
In the loop variable formalism, a gauge transformation is obtained in
the form $\sum _n \la _n \dsn (\Sigma A) = \sum _n \la _n (\dsn
\Sigma) A + \Sigma \dsn A)$. Thus ${d\over d \Sigma}((\dsn \Sigma ) A +
\Sigma \dsn A) =0$ after an integration by parts. Thus the equation
of motion obtained by varying w.r.t $\Sigma$ is unchanged by gauge
transformation.)

However the two point integral diverges even when the particles are on shell. 
It has to be cutoff both in the IR and in the UV. It thus acquires a dependence
on $R/a$ and when we operate with the RG operator,  will give something
non zero.  This is unacceptable because the kinetic term in the action
should vanish on shell. Thus it is necessary to divide by this (formally) divergent
 factor before operating with $d\over d~ln~a$.  This argument motivates the following
equation for the space-time action:
\be   \label{Action}
S = {\p \over \p~ln~a} \{{Z \over {\int _{-R+a} ^{R}dz~ \int _{-R}^{z-a} dw {1\over (z-w)^2}}}\}
\ee

Note that the denominator is form invariant under  $SL(2,R)$ transformations. 

The regulated denominator is ${2R\over a} -1 - ln~{2R\over a}$ which is $2R\over a$ for 
$R>>a$. So the equation becomes
\be
S = {\p \over \p ~ln~a} {aZ\over R} = {a\over 2R}(1+{\p \over \p ~ln~a})Z =
{a\over 2R}(1+ \beta ^i {\p \over \p g^i})Z
\ee
where $g^i$ are various coupling constants of the  2-D field theory (spacetime fields of the string) and 
$\beta ^i = -{dg^i \over d~ln~a}$ (and we have used (\ref{beta})) are the beta functions describing the RG flow.

In the last  form  written, it agrees with the expression given in  
Witten's background independent	
formulation  \cite{WiI,WiII,LW}, particularly as used in \cite{ShI,ShII} 
for tachyon backgrounds. For on-shell
backgrounds it is easy to see that it gives the right answer: 
For on shell fields we can certainly take 
$a\rightarrow 0$ and then the partition function gives the S-matrix 
multiplied by a Mobius infinity which has the form ${R\over a}~ln~{R\over a}$. 
Dividing by $R\over a$ and
operating with $d \over d ~ln ~ a$ gives the S-matrix. The regularization 
subtracts the "on-shell" poles. This 
gives the effective action that reproduces the tree level S-matrix.

Our philosophy here is that  we have a prescription that gives the right on-shell answer,
 and is gauge
invariant (after using the loop variable formalism and replacing $d \over d~ln~a$ by $d\over d~\Sigma$)
off-shell also. This is therefore a serious candidate for the gauge invariant off-shell action. 
In this formalism the engineering dimension of the integrated vertex operator $\int {dz\over a}
V(z)$ is $N-1$ where $N$ is the dimension of $V$ and $-1$ for the $a$ in the denominator. 
The extra power of $a$ in
the gauge fixed formalism, has the same effect in the loop variable formalism as the replacement of
$q_0^2 =N$ by $q_0^2=N+1$.  $q_0^2$ is supposed to count all powers of $a$ -
thus it counts powers of $a$ that come from all sources (including the powers of
$z/a$ that come when vertex operators are Taylor expanded) - other than the anomalous dimension $\ko ^2$
of a vertex operator of momentum $\ko$.  $\ko ^2$ in fact vanishes because $\ko$ is the total momentum,
which is zero for the terms in an action because of the integration over space-time.
Thus we can set $q_0^2 = {d\over d ln~a}$.   Thus the extra powers of $a$ coming from
the factor dividing $Z$, will automatically contribute. In fact the more exact form
of the denominator contains powers of $ln~a$ and in this case $q_0^2$ is not an integer,
so in such situations the last form of $q_0^2$ is more appropriate. 

Actually, since the regularization scheme is important, it is more appropriate to give
the prescription for the action without specifying any particular regularization scheme:
\be     \label{action}
S = {\p \over \p ln ~ a} \{ {Z \over \int dz \int dw \langle O(z) O(w) \rangle}\}
\ee
Here $O$ is a dimension-one operator and it is understood that the same scheme
regularization 
is used in the denominator and the numerator.

\section{Tachyon}
\setcounter{equation}{0}

We can calculate some of the leading order terms in the tachyon action just to see how it works.
Let us start with the zero momentum tachyon, because it is the simplest and also gives
the form of the tachyon potential. 

The world sheet action is $\int _{-R}^{+R} {dz \over a} \phi _0 = -{2R\over a} \phi _0$
Thus $Z(\phi _0 ) = Z_0 e^{-{2R\over a} \phi _0}$ and following our prescription, the space-time action is
\be     
S(\phi _0) \approx  (1+ {d\over d~ln~a})Z = 
(1+{2R\over a} \phi _0)e^{-{2R\over a} \phi _0}
\ee
Upto overall normalization. If we redefine ${2R\over a}\phi _0 = T_0$, then the tachyon
potential is proportional to
\be
V(T_0) \approx (1+T_0)e^{-T_0}
\ee

This form of the potential has been noted earlier in the background independent
formalism \cite{ShI,ShII,KMM}. 

We  now consider the tachyon at non-zero momentum. One should
exercise some care in the choice of regulator since the off-shell answer depends
on this choice. We have seen in \cite{BSERG} that the exact RG has a nice form
if one chooses a smooth regulator. However in principle any regulator is allowed
as long as the continuum limit can be taken, because the S-matrix, 
which involves correlators in a conformal field theory,  does not depend
on the regulator. For our purposes we take a simple form of the regulator
namely - we introduce a short distance cutoff in the limits of integration
of the Koba-Nielsen variables \cite{BSPT}. This is simple to calculate with and illustrates
the general idea. (For the gauge invariant calculation that is done in the next section 
this is not possible and one must use a cutoff Green function.)
Thus our prescription gives 
\[
S= {\p \over \p ~ln~a}[({a\over 2R}) \int _{-R+a} ^{R}dz ~\int _{-R} ^{z-a} dw~\int dk ~\int dp
\]
\[
\langle :e^{ik.X(z)}: :e^{ip.X(w)}:\rangle a^{{k^2+p^2\over 2}-2}\phi (k) \phi (p)
\]
\[
= {\p \over \p ~ln~a}[({a\over 2R}) \int _{-R+a} ^{R}dz ~\int _{-R} ^{z-a} dw~\int dk ~\int dp
\delta (k+p) (z-w)^{k.p} a^{{k^2+p^2\over 2}-2}\phi (k) \phi (p)
\]
\[ =
\int dk \phi(k)\phi (-k)
{\p \over \p ~ln~a} \{ ({a\over 2R}) {1\over (1-k^2)}\{ ( {2R\over a} )-
{  [  ({2R\over a})^{-k^2+2}-1]\over -k^2+2} \}
\]
\be
=\int dk \phi(k)\phi (-k) ({a\over 2R})\{ {(k^2-1) ({2R\over a})^{2-k^2} -1\over (k^2-1)(2-k^2)}\}
 \phi (k) \phi (-k)
\ee

Near $k^2=2$ one has to be more careful because of the pole. When expanding in 
powers of $(k^2-2) ln~({2R\over a})$ one has to keep track of the logarithms in
the denominator as well. In that case one obtains (keeping the zero momentum tachyon as well):
\[
{\p \over \p ~ln~a}  \{ e^{-{2R\over a}\phi _0}[1~+~ {\{ {2R\over a} -1 - ln~{2R\over a} ~-~
({2-k^2\over 2}) ln ^2~{2R\over a} \} \over {2R\over a} -1 -ln~{2R\over a}}\phi  (k) \phi (-k) ] \} 
\]
\[ \approx
{\p \over \p ~ln~a}  \{ e^{-{2R\over a}\phi _0} [{1\over {2R\over a} -1 -ln~{2R\over a}}
+ (1 - {a\over 4R} (2-k^2) ln^2 ~{2R\over a} ) \phi (k) \phi (-k) ] \}
\] 
\[
\approx
(1+ {2R\over a} \phi _0) e^{-{2R\over a}\phi _0} + ({2-k^2\over 2})[ln^2 ~ {2R\over a} +
2~ln~{2R\over a} ]\phi (k) \phi (-k) e^{-{2R\over a}\phi _0}
\]
where we have assumed that ${R\over a} >> ln {R\over a} ,1$ and in the last step
we have also dropped an overall normalization. $R\over a$ is a free parameter
that doesn't affect on-shell physics, but does modify the off-shell action.

One can also calculate the cubic term for on-shell tachyons and one gets in the numerator
an expression similar to the denominator except for a factor of $ln~(z-w)$:
\[
\int  _{-R+a}^{R}  dz~\int  _{-R}^{z-a}  dw~\int _{w+a} ^{z-a} dv {1\over z-w}{1\over z-v}{1\over v-w}
\]
\[ \approx
\int _{-R}^{R+a} dz~\int _{-R}^{z-a} dw {ln~({z-w\over a})\over (z-w)^2} 
\]
\be
= {2R\over a} -1 - ln~({2R\over a}) -\hf ln^2 ~({2R\over a}) 
\ee
The details of the cutoffs in the limits of integration are important. Thus we have chosen
the limits in a simple way - making sure that every integral is cutoff by the same distance
$a$.  This has the advantage that when we apply our prescription the coefficient of
the cubic term is
\be
{\p \over \p ~ln~a} \{ {2R\over a} -1 - ln~{2R\over a} - \hf ln^2 ({2R\over a})
\approx {a\over 2R} \hf [ln^2~({2R\over a}) + 2 ln~({2R\over a})]
\ee
This is exactly the coefficient of the kinetic term. This means that the action 
has the correct relative normalization between the cubic term and the kinetic term, for nearly
on shell tachyons.

\section{Gauge Invariant Formalism}
\setcounter{equation}{0}

We consider the Polyakov action and partition function with some background fields turned on.
 We restrict ourselves to open strings, so that the vertex operators are on the boundary. 
 A review of the loop variable formalism  is contained in \cite{BSREV}. We give a short outline here.
The basic idea, in 
the loop variable formalism, is to define the partition
in terms of two objects, written below as $W[k_i]$ and $\Psi [k_i; \phi ]$.  $\Psi$ is the 
equivalent of a "wave function" that has information about the background fields,
genericall denoted by $\phi_j$ here. $k_i$ are generalized momenta used in the 
Fourier transformation from $X(z)$ to momentum variables. They are conjugate
to the various vertex operators. For the free case $\Psi$ would be free fields and can also 
be thought of as the wave functions in the first quantized picture, expressed in momentum 
space. For example the usual expression $\phi [X(z)] = \int d k_0 e^{ik_0 X (z)} \phi (k_0)$ 
for the tachyon,  can be thought of as a wave function and, after second quantization, a field.
This is generalized to all the states of the open string:
\be
\phi [X] + A_\mu [X] \p X^\mu + S_{\mu \nu} \p X^\mu \p X^\nu +...
= \int \prod _n d\kn \gvkt \Psi [\kn , \phi , A_\mu ,...]
\ee 
where $\tY_n = {\p ^n X \over (n-1)!}$ for $n >0$ and $\tY _0 =X$. 
However for the interacting case $\Psi$ is a generalization of the idea
of a field, because it has  the field  as well as products of the fields at different
spacetime points. (see \cite{BSREV}). In terms of $W$ and $\Psi$, $Z$ is given by:

\[
Z[\phi ]=\int \prod _i {\cal D} k_i (z)\int {\cal D}X \int {\cal D }\al 
~e^{i\sum _{n,m}\int dz_1 \int dz_2 {\bar \kn (z_1)}.
{\bar \km (z_2)} [G + \Sigma ]_{n,m} }\Psi [k_i (z); \phi ]
\]
\be
= \int \prod _i {\cal D} k_i (z) W[k_i(z) ]\Psi [k_i (z), \phi ] 
\ee

All the information
regarding the structure of the equations of motion (EOM) is contained in $W$. 
The information regarding the particular background is contained in $\Psi$.

In the interacting loop variable formalism we rewrite 
\[
\sum _{n\ge 0}\kn (z)\tY _n(z)
\]
\be   \label{Taylor}
=\sum _{n \ge 0}\bar \kn (z) \tY_n (0)
\ee
 by a Taylor expansion of $\tY _n(z)$ about $z=0$.  This defines the 
$\bar \kn (z)$as a function of $\kn (z) $and $z$:
\be
\bar k _q (z) = \sum _{n=0}^q \kn (z) D_q^n z^{q-n}
\ee
with $D^q_q=1, D^n_q = ^{q-1}C_{n-1} , D^0_q={1\over q}$. (Our notation is a little
different from \cite{BSREV} : $D^n_q$ vs. $D^q_n$.)
This defines all the loop variables at one point $z=0$ and we get
$e^{i\int dz \sum _{n \ge 0}\kn (z) \tY _n(z) } = e^{i\int dz \sum _{n \ge 0}\bar \kn (z) \tY_n(0)}$. 
This looks like
a free theory with generalized momenta $\kn$ replaced by $\int dz \bar \kn (z)$. This
can then be covariantized using $\aln$ just as in the free theory and we get a
loop variable
\be
e^{i\int dz \sum _{n\ge 0}\bar \kn (z) \yn (0)}
\ee 
It is very easy to write down interacting gauge invariant equations of motion
because the theory looks exactly like a free theory and the same tehniques
can be used. The gauge transformations are exactly the same as
in the free theory:
\be
\int dz ~\bar \kn (z) \rightarrow \int dz ~\bar \kn (z) + \int dz' ~\la _p (z') \int dz ~\bar k_{n-p} (z)
\ee

We illustrate below an example of a calculation using this formalism. We take the simplest example
involving gauge invariance, which is the massless vector, and derive Maxwell's equation.

This is by far the simplest case since it involves only terms with $\ki$. Let us consider the EOM
for the vector in the loop variable formalism:
\[
{\delta \over \delta \Sigma} \{(\sum _{n=1}^\infty \int dz_1\int dz_2~
\bar \kn (z_1). \ko (z_2) {\p \Sigma \over \p \xn}
)~e^{i \sum _m \int dz_3~\bar \km (z_3) \ym (0)}\]
\be  \label{II}
+ (\int dz_1\int dz_2~
\ko (z_1).\ko (z_2) \Sigma )~e^{i\sum _m \int dz_3~\bar \km (z_3) \ym (0)} \} =0
\ee
And we are instructed to pick the coefficient of the vertex operator $\yim e^{ik.Y} $
 which, on setting $\xn = 0$ is $ \p_z X^\mu \e$.

\[ \Rightarrow
\int dz_1\int dz_2~\{-\sum _{n=1}^\infty \bar \kn (z_1). \ko (z_2) 
i\int dz_3 \sum _{m\ge 0} \bar \km (z_3) Y_{m+n} (0)
e^{i \sum _m \int dz~\bar \km (z) \ym (0)} +\]
\[
\ko (z_1).\ko (z_2)e^{i \sum _m \int dz~\bar \km (z) \ym (0)}\}
\]

We want the coefficient of $\tY _1 (0)$, and this is clearly:
\[
[- \bar \ki (z_1) . \ko (z_2) i \ko (z_3)  + \ko (z_1). \ko (z_2) i \bar \ki (z_3)]\tY _1(0)e^{i\int dz \ko X(0)} 
\]
 
Using the expansion for $\bar \ki (z)$ we see Maxwell's equations emrge.

The prescription for the action is as before
\be
S = {\p \over \p \Sigma} {Z\over \int dz \int dw \langle O(z) O(w) \rangle }
\ee
 The leading term when we differentiate wrt $\Sigma$ is just $k_0^2 + q_0^2$. $k_0^2=0$ 
because the total momentum of all the fields adds up to zero- as explained earlier. Thus we get
in the leading term ${\p \over \p \Sigma} = q_0^2={d\over d~ln~a}$ which is essentially
the same prescription that we had in the gauge fixed case. Thus the leading term is the 
same (provided we do a resummation of the Taylor expansion that is done in this formalism)
 and the other terms are required to implement gauge invariance. Gauge invariance
(as explained in Section 2) follows from the fact that we have a derivative w.r.t $\Sigma$ -
in the loop variable formalism this guarantees gauge invariance.  Thus we have an expression that
is gauge invariant off-shell and also gives for the gauge fixed part (that multiplies $q_0^2$)
the action of Section 2. This is thus our candidate for the gauge invariant space-time action.
In the next section we discuss the resummation.

\section{Resummation}
\setcounter{equation}{0}

In the last section we obtained a gauge invariant action. Both, this action, as well as
the equations of motion derived in earlier papers, suffer from a defect: 
 the pole structure of the Veneziano amplitude is not visible 
unless one includes the effect of all the $\bar \kn$. This can be traced to the Taylor expansion
about one point. It is as if the propagator ${1\over p^2+m^2} $has been written out
as a power series :${1\over m^2} -{p^2\over m^4} +...$ and all the contributions from
different mass particle propagators are reorganized in powers of $p^2$. It is not a problem of principle
but the pole structure is useful to determine the spectrum etc. The Taylor expansion was necessary for
making the full gauge invariance manifest. Having obtained a gauge invariant expression
it is useful, if possible to perform a resummation so that
 we have an expression in terms of $\kn$ rather than $\bar \kn$. In this section we show 
that a resummation can be done.  The part that would be there in the gauge fixed
action is easy to resum. The rest of the terms that are necessary only for gauge invariance
are a little more difficult to resum. 
Nevertheless an expression in terms of Laplace transforms can be written
down.  We describe this method in this section. Explicit computations
of terms in the action are left for the future.

Let us look at the kinds of terms that occur in the eqautions of motion
or action.

When we vary wrt $\Sigma$ in $(k_0^2 + q_0^2)\Sigma $ we get in the action 
some terms that involve the above loop variable
in the form 

\[ 
\sum _{n,m \ge 0}\int dz_1 \int dz_2 \bar \kn (z_1) \bar \km (z_2) G_{n,m} (0) 
\]
\[
= \sum _{n,m \ge 0}\int dz_1 \int dz_2 \bar \kn (z_1) \bar \km (z_2)\langle \yn (0) \ym (0) \rangle 
\]
In this form one does not see the logarithmic form of the two point function, because
a Taylor expansion has been performed:
\[
ln~ [(z_1-z_2)^2 + a^2 ] = ln~ a^2 + {z_1^2\over a^2} + {z_2 ^2\over a^2} - {2z_1 z_2 \over a^2}+.....
\]
and the $z_1, z_2$ dependences are all in the $\bar \kn (z_i)$.  We can easily resum and undo this
by first going to a gauge where $\xn =0$, so that $ \yn =\tY _n$ and then resum using (\ref{Taylor}), to get
\be
\sum _{n,m \ge 0}
\int dz_1 \int dz_2  \kn (z_1) \km (z_2) G_{n,m} (z_1,z_2) 
\ee

where 
\[ 
G(z_1, z_2) = \langle Y (z_1) Y(z_2)\rangle |_{\xn =0} = \langle X(z_1) X(z_2) \rangle 
\]

 In this form the equations involve only $\kn$ and furthermore the pole structure
is easily seen to emerge on integrating over the $z_i$.

The complication starts when one considers the effect of varying
$\Sigma$ in $(\kn . \ko + q_n q_0)\dsn \Sigma$. We get, on integrating
by parts, terms of the form
\[
= \sum _{n,m \ge 0}\int dz_1 \int dz_2 \bar \kn (z_1) \bar \km (z_2)
\langle   ({\p \over \p x_p} \yn (0)) \ym (0) \rangle 
\]

We thus need to be able to do sums of the form
\be
\sum _{n\ge 0} \bar \kn (z) \tY _{n+m} (0)
\ee

Our strategy will be to relate this by some mathematical operation to 
$\sum _{n\ge 0} \bar \kn (z) \tY _{n} (0) $ whose sum we know. To this end observe
that
\[
\bar \kn (z) = \kn + (n-1) k_{n-1} z + {(n-1)(n-2)\over 2!} k_{n-2}z^2 + ...+  \ki z^{n-1} + {\ko \over n} z^n
\]
\[
\p _z \bar \kn (z) = (n-1) [k_{n-1}  + {(n-2)} k_{n-2}z + ...+  \ki z^{n-2}  + {\ko \over (n-1)}z^{n-1}] = (n-1)\bar k_{n-1} (z)
\]
\[
(\p _z)^2 \bar \kn (z) = (n-1)(n-2) \bar k_{n-2}(z)
\]
\[
(\p _z)^m \bar \kn (z) =
{(n-1)!\over (n-m-1)!}\bar k_{n-m}(z)
\]
Thus we need to perform the sum over $n$ in
\[ I=
\sum _{n=m}^\infty \bar k_{n-m} (z) \tY _n (0) = \sum _{n=m+1} ^{\infty}  (\p _z)^m \bar \kn (z) \tY _n (0) {(n-m-1)!\over (n-1)!}
+  (\p _z )^m \bar \km (z) {\tY _m (0)\over (m-1)!}
\]
We can use Laplace and Inverse Laplace transforms to get the factorials. Let
\[ 
\sum _m a^m F_m = F(a) 
\] 
\[ \int da~ e^{-as} F(a) = {\cal F}(s) = \sum _m {m!\over s^{m+1}} F_m
\]
This enables us to get $m!$ in the numerator. Let
\[
{\cal G}(s) = \sum _m {G_m \over s^{m+1}} 
\]
Then
\[
G(a) = \int _{\gamma -i\infty}^{\gamma +i\infty} {\cal G}(s) e^{as} = 
\sum _m {a^m \over m!} G_m
\]

The contour for the $s$ integral has to be chosen to the right of all the singularities of ${\cal G} (s)$.
This enables us to get $m!$ in the denominator. All we need is to
 introduce appropriate "chemical potentials"
as counters for the relevant indices.
Thus define
\be   \label{Fsa}
{\cal F}_m(s,a) = \sum _{n=m+1} ^{\infty}  {a^{n-m-1}\over s^n}
 (\p _z)^m \bar \kn (z) \tY _n (0) +
 {1\over s^m} (\p _z )^m \bar \km (z) \tY _m (0)
\ee
Define
\be	\label{Ftb}
F_m(t,b) = \int _{\gamma -i\infty}^{\gamma +i\infty} ds~
\int _0^\infty da~ e^{ts} e^{-ba} {\cal F}_m(s,a)
\ee
Then the required answer 
\be	\label{I}
I= F_m(1,1)
\ee

The sum in (\ref{Fsa}) does not have any of the factorials,
 but one has to deal with the factors of $a,s$. 
This can be gotten rid of  by some rescalings. Let 
\be
\kn = \kn ' ({s\over a})^n ~~~; ~~~z= Z({s\over a})
\ee
Then
\be    \label{5.7}
\bar \kn (z) = ({s\over a})^n \bar \kn ' (Z) =  ({s\over a})^n[\kn ' (z) + ...+ 
D_n^m \km ' (z) Z^{n-m} +...+ \ko {Z^n\over n}]~~~;~~~\p _z = {a\over s} \p _Z
\ee

\[ {\cal F}_m(s,a) = a^{m-1}\sum _{n=m+1}^\infty \p _z^m \bar \kn ' (Z) \tY_n (0) 
+ a^{-m} \p _z^m \bar \km ' (Z) \tY _m (0)
\]
Now the sum on $n$ can be done:
\[ \sum _{n=m+1} ^\infty \bar \kn ' (Z) \tY _n (0) = 
\sum _{n=0}^\infty \bar \kn ' (Z) \tY _n (0) - \sum _{n=0}^m \bar \kn ' (Z) \tY _n(0)
\]
\[ 
= \sum _{n=0}^\infty  \kn ' (z) \tY _n (Z) - \sum _{n=0}^m\bar \kn ' (Z) \tY _n(0)
\]
The second term can also be written in terms of $\kn ' (z)$ 
but since only a finite number of terms are involved it doesn't make
much difference. Finally we can also go back to $\kn$:
\[ 
{\cal F}_m(s,a) = {1\over a^m}\{{1\over a}[ \sum _{n=0} ^\infty \kn (Z) \p _z ^m \tY_n (Z)({a\over s})^n 
- \sum _{n=0}^m ({a\over s})^n \p _z^m \bar \kn (Z) \tY_n (0)] +
 ({a\over s})^m\p_z^m \bar \km (Z) \tY_m(0)\}
\]
\[=
 {1\over a^m}\{{1\over a}[ \sum _{n=0} ^\infty \kn (Z) \p _z ^m \tY_n (Z)({a\over s})^n 
-  ({a\over s})^m \p _z^m \bar \km (z) \tY_m (0)] +
 ({a\over s})^m\p_z^m \bar \km (z) \tY_m(0)\}
\]
\[=
 {1\over a^m}\{{1\over a}[ \sum _{n=0} ^\infty \kn (Z) \p _Z ^m \tY_n (Z)({a\over s})^{n+m} 
-  ({a\over s})^m  \ko \p _z^mX(0)] +
 ({a\over s})^m\ko \p_z^m  X(0)\}
\]
Writing the $n=0$ terms separately gives finally:
\be
{\cal F}_m(s,a;z)=
\sum_{n=1}^\infty {a^{n-1}\over s^{n+m}}\kn (z) {\p_Z^{n+m} X(Z)\over (n-1)!} +
 {1\over as^m}\ko [ \p_Z^m X(Z) - \p_Z^mX(0)]+{1\over s^m}\ko \p_z^mX(0)
\ee

We have specified the location $z$ to be complete. This is to be inserted inside correlation functions
and so becomes a function of $s,a,Z$ or equivalently $s,a,{a\over s}z$. One has to do the integrals over
$s,a$ in addition to the usual integration over vertex operator location, $z$. This is the price one pays
for obtaining an answer in terms of $\kn$.
Thus the final answer is
\be
\sum _{n=m}^\infty \bar k_{n-m} (z) \tY _n (0) = F_m(1,1)=\int _{\gamma -i\infty}^{\gamma +i\infty} ds~
\int _0^\infty da~ e^{s} e^{-a} {\cal F}_m(s,a;z)
\ee

In addition to the above, the loop variable expression for the action or equations of motion,
 involves terms of the form
$\sum _{n,m\ge 0}\bar \kn  ^\mu (z_1) \bar \km  ^\nu (z_2) \tY ^\rho _{n+m}(0)$ and also 
$\sum _{n,m,p \ge 0}\bar \kn ^\mu (z_1)\bar \km ^\nu (z_2)\bar k_p ^\rho (z_3) \tY ^\rho _{n+m+p}(0)$.
The resummed versions of these are derived in the Appendix.  The results are:
\[
\sum _{n,m\ge 0}\bar \kn  ^\mu (z_1) \bar \km  ^\nu (z_2) \tY ^\rho _{n+m}(0)=
\]
\[
\int d\mu ~ \Big( \sum _{n\ge 0}({a_1\over s})^n \kn ^\mu(z_1) {\p_z ^n \over (n-1)!} + (a_1-1)\kom (z_1) \Big)
\Big( \sum _{m\ge 0}({a_2\over s})^m \km ^\nu (z_2) {\p_z ^m \over (m-1)!} + (a_2-1)\kon (z_2) \Big)
\]
\be     \label{rsum2}
X^\rho (z+Z_1+Z_2)|_{z=0}
\ee
where 
\[
\int d\mu \equiv \int _{\gamma -i\infty}^{\gamma +i\infty} ds ~e^{s} 
\int _0^\infty {da_1\over a_1}~e^{-a_1}
\int _0^\infty {da_2\over a_2}~e^{-a_2}
\]
In using the above formula the following rule must be applied: 
When a factor $(a_1-1)$ occurs, then the argument of $X$is
$(z+Z_2)$ and when $(a_2-1)$ is used the argument is $X(z+Z_1)$ 
and when both factors occur, the argument is $z$. Otherwise the 
argument is as indicated: $z+Z_1+Z_2$. Thus for instance one term would be
$(a_1-1) \kom (z_1) (a_2-1) \kon (z_2) X^\rho (0)$.
It is also understood that in the above expressions, when $n=0$,
 the factors of $(n-1)!$ that occur in the denominators 
have to be replaced by $1$.

Similarly 
\[
\sum _{n,m,p \ge 0}\bar \kn ^\mu (z_1)\bar \km ^\nu (z_2)\bar k_p ^\rho (z_3) 
\tY ^\sigma _{n+m+p}(0) =
\]
\[
\int d\mu ~ \Big( \sum _{n\ge 0}({a_1\over s})^n \kn ^\mu (z_1) {\p_z ^n \over (n-1)!} +
 (a_1-1)\kom (z_1) \Big)
\Big( \sum _{m\ge 0}({a_2\over s})^m \km ^\nu (z_2) {\p_z ^m \over (m-1)!} +
 (a_2-1)\kon (z_2) \Big)
\]
\[
\Big( \sum _{p\ge 0}({a_3\over s})^p k_p^\rho(z_3) {\p_z ^p \over (p-1)!} + 
(a_3-1) \ko ^\rho  (z_3) \Big)
\]
\be	\label{rsum3}
X^\sigma (z+Z_1+Z_2+Z_3)
\ee

with
\[
\int d\mu \equiv \int _{\gamma -i\infty}^{\gamma +i\infty} ds ~e^{s} 
\int _0^\infty {da_1\over a_1}~e^{-a_1}
\int _0^\infty {da_2\over a_2}~e^{-a_2}
\int _0^\infty {da_3\over a_3}~e^{-a_3}
\]

It is understood that in the above expressions, when $n=0$,
 the factors of $(n-1)!$ that occur in the denominators 
have to be replaced by $1$.
Also as above, when the factor $(a_i-1)$ occurs the corresponding $Z_i$ is dropped from the argument of $X$.

This resummation can be applied to the equations of motion derived in earlier papers, or to the action
described in this paper, though this is not done in this paper.

\section{Conclusions}

We have presented a candidate gauge invariant action within the loop variable formalism.
It was constructed as an off shell gauge invariant generalization of a formula, viz. (\ref{action}), that
gave the right kinetic term and the cubic interaction for the tachyon.
 We also gave a general argument that this same formula should
give the effective action that reproduces the S-matrix. 
We checked our prescription for the tachyon and it gives results for the potential that
is very similar to that obtained earlier in Witten's formalism.
The gauge invariance is obtained by replacing
$d\over d~ln~a$ by $d\over d\Sigma$ and using the loop variable form for the world sheet action.
 We believe that because
 space-time gauge invariance is built into this method in a way that does not rely  on world-sheet 
reparametrization invariance or BRST invariance, this formalism can also be  
manifestly background independent.
 
In the loop variable formalism a Taylor expansion is made in order to obtain
gauge invariant equations. In the process one loses the pole structure
of the string amplitudes.  In this paper we have given a method of resumming the terms.
In the "physical" part of the action (i.e. the part that contributes to the S-matrix of physical
particles) this is easy to do and the pole structure can be made manifest.
 For the remainder of the terms also (i.e. those necessary for
gauge invariance) we have given a procedure that does the sum and the answer involves doing
Laplace transform integrals. 

There are many open questions that remain. While we have a gauge invariant formalism, we have
only used it in this paper for the tachyon, where there are no issues of gauge invariance. We need
to work out the actions for some of the higher spin fields. The resummation techniques 
given in this paper should be useful for this. Then there is always the question of closed
strings, and also of extending this to curved space. We hope to return to these issues soon.

\appendix
\section{Appendix}
\setcounter{equation}{0}

We derive (\ref{rsum3}) here. We need to evaluate
\[
III=\sum _{n,m,r=0}^{\infty} \bar \kn (z_1) \bar \km (z_2) \bar k_r (z_3) \tY _{n+m+r}(0) 
\]
which we rewrite as
\[
\sum _{n,m,r=0}^\infty \bar \kn (z_1) \p _z ^n ~~\bar \km (z_2) \p _z ^m ~~\bar k_r (z_3) {\p _z ^r X(z) \over (n+m+r-1)!}|_{z=0 }
\]
\[ = \sum _{n=0}^\infty (n-1)! \bar \kn (z_1) {\p _z^n \over (n-1)!} \sum _{m=0}^\infty (m-1)! \bar \km (z_2) {\p _z ^m \over (m-1)!}
\]\[
\sum _{r=0}^\infty {(r-1)!\over (n+m+r-1)!} \bar k_r (z_3) {\p _z^r \over (r-1)!}X(z) |_{z=0}
\]

(Here and below, $(n-1)!$ is to be replaced by $1$ whenever $n=0,1$.)

We have to get rid of the unwanted factorials by using Laplace transforms, so that we can use (\ref{Taylor}).
Thus we get
\[\ \int _{\gamma -i\infty}^{\gamma +i\infty} ds~\int da_1~\int da_2~\int da_3~ e^{s-a_1-a_2-a_3} \Big( \sum _{n\ge 1}^\infty a_1^{n-1}\bar \kn (z_1) {\p _z ^n \over (n-1)!}
+ \bar \ko (z_1) \Big)
\] 
\[\Big( \sum _{m\ge 1}^\infty a_2^{m-1}\bar \km (z_2) {\p _z ^m \over (m-1)!}
+ \bar \ko (z_2) \Big)
\Big( \sum _{r\ge 1}^\infty a_3^{r-1}\bar k _r (z_3) {\p _z ^r \over (r-1)!}
+ \bar \ko (z_3) \Big) X(z) |_{z=0} {1\over s^{n+m+r}}
\]
\[\ =
\underbrace {\int _{\gamma -i\infty}^{\gamma +i\infty} ds~\int {da_1\over a_1}~\int {da_2\over a_2}~\int {da_3\over a_3}~ e^{s-a_1-a_2-a_3} }_{=\int d\mu}\Big( \sum _{n\ge 1}^\infty ({a_1\over s})^n\bar \kn (z_1) {\p _z ^n \over (n-1)!}
+ a_1\bar \ko (z_1) \Big)
\] 
\be   \label{III}
\Big( \sum _{m\ge 1}^\infty ({a_2\over s})^m\bar \km (z_2) {\p _z ^m \over (m-1)!}
+ a_2\bar \ko (z_2) \Big)
\Big( \sum _{r\ge 1}^\infty ({a_3\over s})^r\bar k _r (z_3) {\p _z ^r \over (r-1)!}
+ a_3\bar \ko (z_3) \Big) X(z) |_{z=0} 
\ee
We now rescale to get rid of the factors $({a_i\over s})$:
\[
\kn (z_i) = \kn ' (z_i) ({s\over a_i})^n ~~~~; ~~~z_i = {s\over a _i} Z_i  ~~~~\Rightarrow ~~~\bar \kn (z_i) = ({s\over a_i})^n \bar \kn '(Z_i)
\] 
This means 
\[
\Big( \sum _{n\ge 1}^\infty ({a_1\over s})^n\bar \kn (z_1) {\p _z ^n \over (n-1)!}
+ a_1\bar \ko (z_1) \Big) =\Big( \sum _{n\ge 1}^\infty \bar \kn '(Z_1) {\p _z ^n \over (n-1)!}
+ a_1\bar \ko (z_1) \Big)
\]
\be   \label{a}
 = \Big( [\bar \ko (z_1) +\sum _{n\ge 1}^\infty \bar \kn '(Z_1) {\p _z ^n \over (n-1)!}]
+ (a_1 -1)\bar \ko (z_1) \Big) 
\ee

where $\bar \kn '(Z)$ is defined in (\ref{5.7}).
The expression in square brackets is what occurs in (\ref{Taylor}). (Note that $\bar \ko (z) = \ko (z)$). 
Thus we can write more generally
\be  \label{Taylorprime}
 [\bar \ko (z_1) +\sum _{n\ge 1}^\infty \bar \kn '(Z_1) {\p _z ^n \over (n-1)!}]f(z) = [\ko (z_1) + \sum _{n\ge 1} ^\infty 
\kn '(z_1) {\p _z ^n \over (n-1)!} ]
f(z+Z_1) 
\ee
We can iterate this equation and get for instance:
\[
[\bar \ko (z_1) +\sum _{n\ge 1}^\infty \bar \kn '(Z_1) {\p _z ^n \over (n-1)!}][\bar \ko (z_2) +\sum _{m\ge 1}^\infty \bar \km '(Z_2) {\p _z ^n \over (m-1)!}]f(z) =
\]
\[[\bar \ko (z_1) +\sum _{n\ge 1}^\infty \bar \kn '(Z_1) {\p _z ^n \over (n-1)!}]
\underbrace{[\ko (z_2) + \sum _{m\ge 1} ^\infty 
\kn '(z_2) {\p _z ^n \over (m-1)!} ]f(z+Z_2)}_{g(z+Z_2)}
\]
\[ =
 [\ko (z_1) + \sum _{n\ge 1} ^\infty 
\kn '(z_1) {\p _z ^n \over (n-1)!} ]g(z+Z_1+Z_2) 
\]
\[ =
 [\ko (z_1) + \sum _{n\ge 1} ^\infty 
\kn '(z_1) {\p _z ^n \over (n-1)!} ][\ko (z_2) + \sum _{m\ge 1} ^\infty 
\kn '(z_2) {\p _z ^n \over (m-1)!} ]f(z+Z_1+Z_2)
\]

Now we apply the rescalings, as well as equation (\ref{a}) for each of the factors in (\ref{III}) and apply (\ref{Taylorprime})
repeatedly to get:
\[ III=
\int d\mu~ \Big( [\ko (z_1) + \sum _{n\ge 1} ^\infty 
\kn '(z_1) {\p _z ^n \over (n-1)!} ] + (a_1-1)\ko (z_1) \Big) \]
\[
\Big( [\ko (z_2) + \sum _{m\ge 1} ^\infty 
\km '(z_2) {\p _z ^m \over (m-1)!} ] + (a_2-1)\ko (z_2) \Big)
\]
\[
\Big( [\ko (z_3) + \sum _{r\ge 1} ^\infty 
\kn '(z_3) {\p _z ^r \over (r-1)!} ] + (a_3-1)\ko (z_3) \Big) X(z+Z_1+Z_2+Z_3)|_{z=0}
\]

It must be kept in mind while using the above formula, that when the term $(a_1-1)\ko (z_1)$
 is used in place of the term within
square brackets, $X(z)$  is to be evaluated at $z=Z_2+Z_3$, i.e. $Z_1$ is dropped. 
Similarly if $(a_1-1)\ko (z_1) (a_2-1) \ko (z_2)$
is used, then we evaluate at $z=Z_3$ and so on. This concludes the derivation of (\ref{rsum3}). 
Equation (\ref{rsum2}) is a simpler case of the above where only two iterations are required.

\end{document}